\documentclass[a4paper]{jpconf}

\usepackage{graphicx}
\usepackage{xspace}

\usepackage{aasjournals}
\usepackage{amsmath}
\usepackage{amssymb}
\usepackage{hyperref}
\usepackage[sort&compress,numbers]{natbib}

\usepackage{color}
\newcommand{\msun}{$M_\odot$\xspace}

\newcommand{\unitspace}{\ensuremath{\,}}
\newcommand{\usp}{\unitspace}

\newcommand{\unitstyle}[1]{\ensuremath{\mathrm{#1}}}
\newcommand{\power}[2]{\ensuremath{{#1}^{#2}}}

\newcommand{\centi}{\unitstyle{c}}
\newcommand{\kilo}{\unitstyle{k}}

\newcommand{\meter}{\unitstyle{m}}

\newcommand{\second}{\unitstyle{s}}

\newcommand{\cm}{\centi\meter}
\newcommand{\gram}{\unitstyle{g}}

\newcommand{\grampercc}{\gram\usp\power{\cm}{-3}} 

\newcommand{\erg}{\unitstyle{ergs}} 

\newcommand{\ergspersecond}{\erg\unitspace\power{\second}{-1}}

\newcommand{\Msun}{\ensuremath{M_\odot}}

\newcommand{\km}{\kilo\meter}   

\newcommand{\dwarf}{{\sffamily DWARF}}

\newcommand{\maestro}{{\sffamily Maestro}}

\newcommand{\pynucastro}{{\sffamily pynucastro}}

\newcommand{\yt}{{\sffamily yt}}

\newcommand{\mrm}[1]{\ensuremath{\mathrm{#1}}}
\newcommand{\iso}[2]{\mrm{^{#2}{#1}}}
\newcommand{\xiso}[2]{\mrm{X(^{#2}{#1})}}
\newcommand{\dxiso}[2]{\mrm{\dot{X}(^{#2}{#1})}}

\newcommand{\gcc}{{\mrm{g~cm^{-3}}}}

\begin{document}

\title{Thermonuclear (Type Ia) Supernovae and Progenitor Evolution}

\author{A~C~Calder$^{1,2}$, 
D~E~Willcox$^{1,3}$, C~J~DeGrendele$^1$, D~Shangase$^{1,4}$, M~Zingale$^1$, and D~M~Townsley$^5$}

\address{$^1$ Department of Physics and Astronomy, 
Stony Brook University, Stony Brook, NY 11794-3800, USA}
\address{$^2$ Institute for Advanced Computational Science,
Stony Brook University, Stony Brook, NY 11794-5250, USA}
\address{$^3$ Center for Computational Sciences and Engineering, Lawrence Berkeley National Lab, Berkeley, CA 94720 USA}
\address{$^4$ Department of Physics, University of Michigan, Ann Arbor, MI 48109-1040, USA}
\address{$^5$ Department of Physics and Astronomy, University of Alabama, Tuscaloosa, AL 35487-0324, USA}

\ead{alan.calder@stonybrook.edu}

\begin{abstract}
Thermonuclear (type Ia) supernovae are bright stellar explosions
with the unique property that the light curves can be standardized,
allowing them to be used as distance indicators for cosmological studies.
Many fundamental questions bout these events remain, however. We provide
a critique of our present understanding of these and present results of
simulations assuming the single-degenerate progenitor model consisting of
a white dwarf that has gained mass from a stellar companion. We present
results from full three-dimensional simulations of convection with weak
reactions comprising the A=23 Urca process in the progenitor white dwarf.
\end{abstract}

\section{Introduction}

Thermonuclear (type Ia) supernovae are bright stellar explosions thought
to occur when approximately one solar mass of stellar material composed
principally of C and O burns under degenerate conditions. The light
curves of these events have the unique property that the rate of 
decline from maximum light is inversely proportional to the 
overall brightness. This ``brighter is broader" relationship allows
these events to be standardized and used as distance 
indicators~\cite{phillips:absolute}. Despite
highly successful use of these for cosmology~\cite{riess.filippenko.ea:observational,perlmutter.aldering.ea:measurements,leibundgut2001}, the setting
of these events is still subject to debate. 

At present, there are three progenitor systems widely under study:
the ``single degenerate" scenario in which a C/O white dwarf (WD) 
gains mass from a companion and explodes as it approaches a limiting
mass, the ``double degenerate" scenario in which two WDs merge and
subsequently explode, and the ``double detonation" model in which
a detonation in an accreted He layer triggers a detonation in the
underlying C/O WD~\cite[and references therein]{calderetal2013,maoz:2014,seitenzahltownsley2017,roepkesim2018}.
In this work, we explore the evolution of the progenitor white dwarf
as it approaches the limiting mass in the single degenerate scenario.

\section{The Effects of the History and Composition of the WD Progenitor}

The capability of modern computing allows development of models with
a vast amount of included physics and thus simulations of thermonuclear
supernovae with unprecedented realism. Correspondingly, advances in observational
technology allow a synthesis between observation and theory that allows
study of the connections between progenitors and the observed light
curves and spectra \cite{hoeflichetal2013,hillebrandtetal2013}.
Both approaches allow studying the efficacy of proposed progenitor
systems, and within a given progenitor system allow the study of
systematic effects on the brightness of an event due to age and composition. 

Considerable research on systematic effects has been performed in the
single degenerate paradigm. Research has addressed how properties of
the progenitor WD that should follow from
the age and metallicity of the host galaxy influence the outcome of explosions, including 
the composition of progenitor white dwarf \cite{jacketal2010,bravoetaldiff2011,ohlmannetal2014,milesetal2016,leungnomoto2018} and the internal structure of the progenitor white dwarf \cite{Krueger2010On-Variations-o,bravoetaldiff2011,kruegetal12,seitenzahletal11}. Also, as we will illustrate immediately below, research has
applied the latest in stellar evolution theory to produce new types of progenitors 
and explored the results of explosions occurring in these new progenitors \cite{kromeretal2015,bravoetal2016,willcoxetal2016}. 

As an example of contemporary research, we present highlights from a study of explosions 
from a new class of progenitors. These progenitors arise
from recent results in stellar evolution indicating that convection in the interior of a late-time AGB star can strongly influence the nuclear burning. 
Studies suggest that 
convective boundary layers may inhibit burning in regions of the interior
with the result that supernova progenitors can be a ``hybrid" between traditional C/O and O/Ne white dwarfs \cite{denissenkovetal2013,denissenkovetal2015,brooksetal2017}. 
Our group investigated supernova explosions from these progenitors using the 
hybrid C/O/Ne progenitor models of Denissenkov \cite{denissenkovetal2013,denissenkovetal2015} and compared the results to an earlier study of explosions from traditional C/O progenitors~\cite{Krueger2010On-Variations-o,kruegetal12}. 

\begin{figure*}[!ht]
  \begin{minipage}{0.48\textwidth}
    \includegraphics[width=\linewidth]{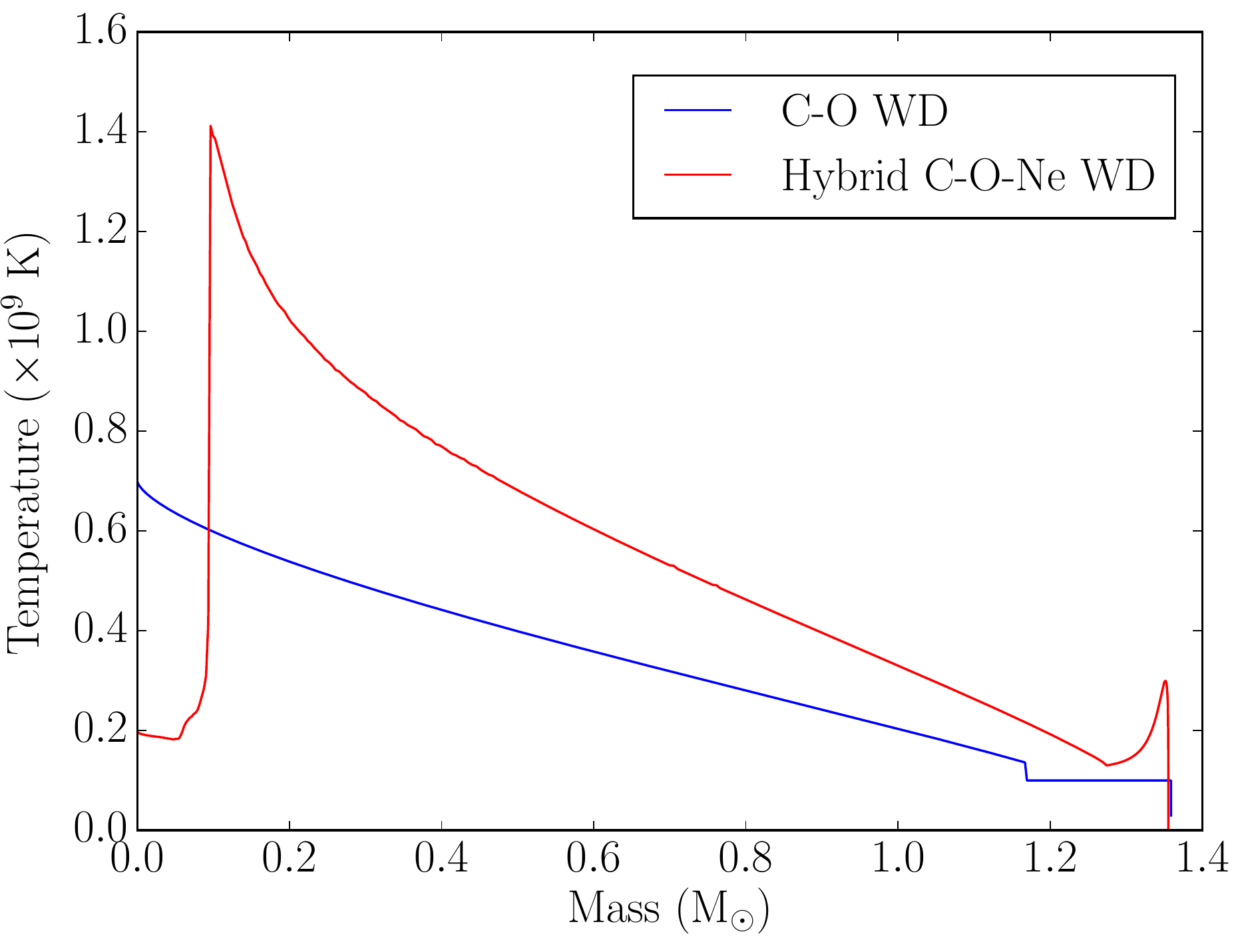}
  \end{minipage} \hfill
  \begin{minipage}{0.48\textwidth}
    \includegraphics[width=\linewidth]{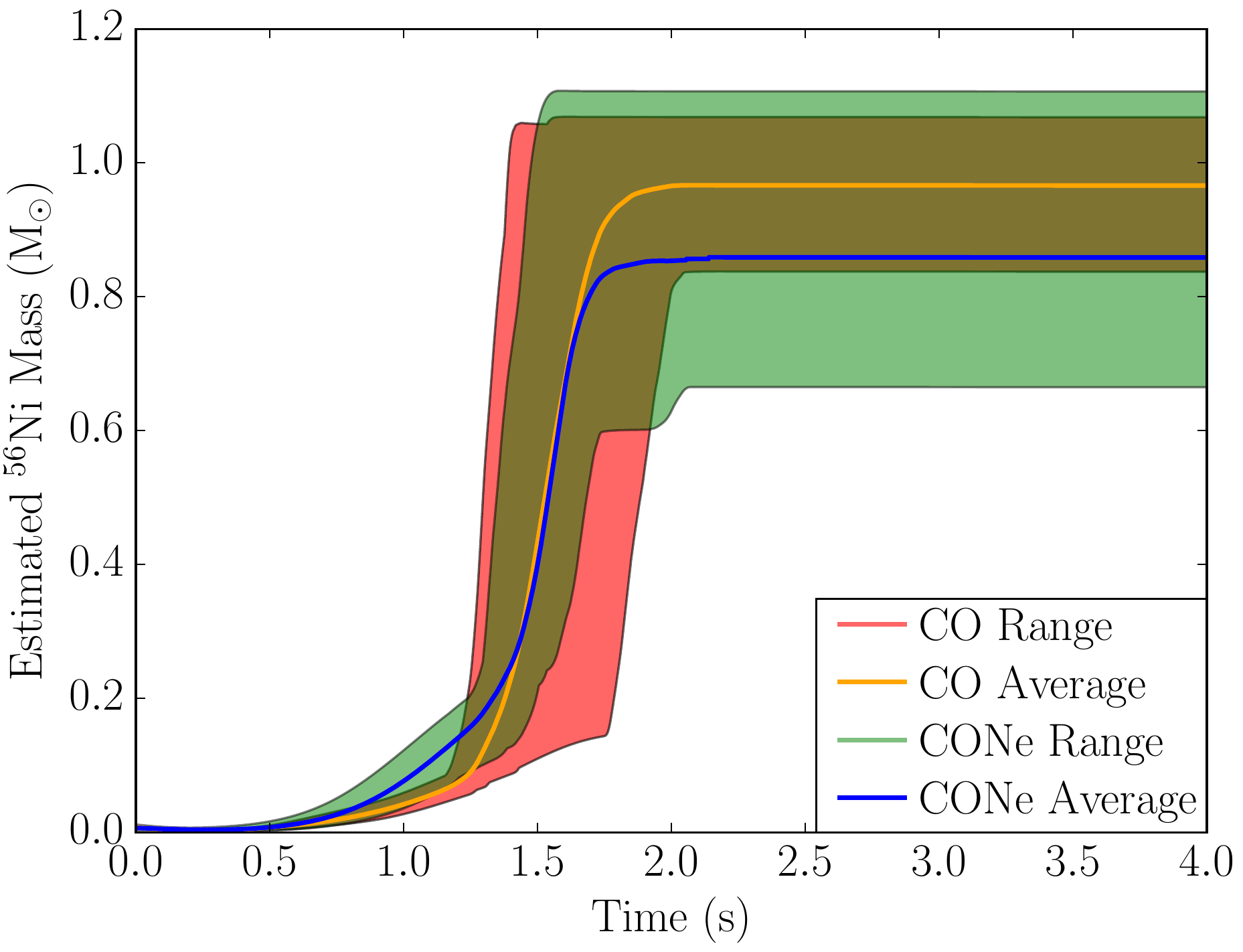}
  \end{minipage} \caption{\label{fig:hybrid} Left panel: Temperature profiles of the hybrid C-O-Ne white dwarf (red) and a reference C-O white dwarf (blue).
Right panel: Evolution of the estimated $^{56}$Ni yields for C-O and Hybrid C-O-Ne WD realizations. The ensemble-averaged value among C-O realizations is shown in yellow, with the full range of values at any point in time for the C-O realizations shown in red. Likewise, the ensemble-averaged value among C-O-Ne realizations is shown in blue and their range of values shown in green. Panels from \cite{willcoxetal2016} \copyright\ AAS. Reproduced with permission.}
\end{figure*}

The left panel of Figure \ref{fig:hybrid} shows the initial temperature profiles of the hybrid progenitors and $^{56}$Ni explosion yields. A unique feature of these hybrid progenitors is that the temperature peak is predicted to lie near a mass coordinate of 0.15~\msun\ due to the composition and energy generation structures during the accretion phase. Thus while the overall structure
of these hybrid progenitors is similar to traditional progenitors, the ignition 
occurs farther out in the star and leads to a slightly different way of burning with the inner core even being consumed last in some simulations.
The right panel of Figure \ref{fig:hybrid} presents the evolution of estimated $^{56}$Ni yields from a traditional progenitor and a hybrid progenitor. 
The results of the study indicated that explosions from these 
progenitors are largely 
consistent with explosions from traditional progenitors, but indicated less kinetic energy in ejecta and slightly less $^{56}$Ni is synthesized on average, which suggests slightly dimmer events. Complete results from the study may be found in \cite{willcoxetal2016}.

Motivated by these earlier studies of systematic effects 
on the brightness of explosions, we initiated a study of conditions
in the progenitor as it approaches ignition.  At this point in the evolution, relatively slow fusion reactions occur that change the composition and drive convection. If conditions are
right, reactions can occur and energy can be lost by a process 
known as the convective Urca process. Below we present an initial
look at simulations of this process in the progenitor white dwarf.

\section{The Convective Urca Process}

The Urca process is a nuclear decay cycle of alternating electron capture
and beta decay on the same nucleus with the net effect of producing 
neutrinos. An electron capture 
\begin{equation} 
e^{-} + \left(Z,A\right) \longrightarrow \left(Z-1,A\right) + \nu_e
\end{equation} 
is followed by a beta decay
\begin{equation} 
\left(Z-1,A\right) \longrightarrow \left(Z,A\right) + e^{-} + \bar{\nu}_e \; ,
\end{equation} 
both of which emit a neutrino or an anti-neutrino~\cite{clayton1968}. The rates of these depend
on the local density, temperature, and composition, with electron captures occurring
at densities high enough for the electron energy to exceed the threshold for electron capture and beta decays occurring at densities low enough that there is significant phase space for the emitted electron to occupy.

The convective Urca process, introduced by Paczynski \cite{paczynski_carbon_1972}, occurs in the single degenerate scenario when the WD has
gained enough mass to compress the core and start slow C fusion reactions that 
drive a period of simmering prior to the runaway. Convection carries a warm blob of
material up to lower densities were it cools and beta decays and then descends to
capture electrons and heat. Neutrinos emitted during this cycle escape the
stellar material ($\rho \lesssim 5 \times 10^9~\gcc$) and transport energy away, cooling the star. Several pairs of nuclei can experience the convective Urca process,
but in this study we consider only the \mrm{A=23} pair, \iso{Na}{23}/\iso{Ne}{23}.

Paczynski introduced the Urca process as an energy loss mechanism to 
delay the thermonuclear runaway of the WD. 
Subsequent analytical work by Bruenn 
accounted for heating due to the relaxation of the electron distribution in regions of higher density than the electron capture threshold and argued that the convective
Urca process would not stabilize the WD core against thermonuclear runaway
\cite{bruenn_thermal_1973}.  Subsequent studies using analytic results and one-dimensional stellar evolution models~\cite{couch_effects_1973,couch_thermal_1974,couch_carbon_1975,lazareff_thermodynamics_1975,Iben78a,Iben78b,Iben82} showed that the convective Urca process is complex and stressed
the need for multidimensional simulations. 

Stein and Wheeler performed two-dimensional simulations of the convective Urca process with 
the \dwarf\ implicit hydrodynamics code in order to evaluate the interaction between electron 
capture, $\beta^{-}$-decay, and convection, but the simulations were limited to a wedge of the
star \cite{Stein2006}. 
The study found that under some conditions, the \mrm{A=23} Urca process limited the extent
of the convection zone to the region of the star interior to the density at which the 
electron capture and beta decay rates balance. This density occurs at a particular radius
of the star and defines the Urca shell. The result of Stein and Wheeler was important because
the extent of the convection zone across the Urca shell influences the energy
evolution of the core and it is not possible to constrain with one-dimensional models.

Contemporary studies performed with detailed stellar evolution simulation codes
must make assumptions about the extent of convection into the Urca shell. 
Denissenkov et al.,
for example, address the problem by performing
simulations with different assumptions about the structure of the convection
zone around the \mrm{A=23} Urca shell and compare the results to the accepted 
evolution path (c.f.\ Figure 9) \cite{denissenkovetal2015}.
Considerable work in stellar evolution
included the convective Urca process in one-dimensional calculations 
\cite{Lesaffre2005A-two-stream-fo,denissenkovetal2013,denissenkovetal2015,mr2016,schwab_importance_2017,schwab_exploring_2017}, 
but the restriction of one dimension requires assumptions about turbulent convection and
the interaction with the Urca shell, an inherently three-dimensional phenomenon. The 
results we present here, a preliminary investigation of the convective 
\mrm{A=23} Urca process with full three-dimensional simulations, is meant to better our understanding of this process and hence better characterize the conditions in the simmering 
white dwarf progenitors of a thermonuclear supernovae.

\section{Simulation Instrument}
The simulation instrument we used for this study was the \maestro\ code, which is part of the AMReX Astrophysics Suite of adaptive mesh refinement hydrodynamics codes for reactive astrophysical flows~\cite{amrex}. The hydrodynamics in \maestro\ is designed for low Mach number flows as are found in applications like convection. \maestro\ uses a projection method to filter out acoustic waves but accounts for compressibility effects due to energy generation and gravitational stratification \cite{Almgren2005Low-mach-number, Almgetal06b, almgren_low_2008}. The method thus allows the time step to be set by the advection speed and not the sound speed, which allows for considerably larger time steps than traditional hydrodynamics methods. 

\maestro\ has been applied to a host of astrophysics problems including an earlier study of WD convective
simmering on which this study of the Urca process is based \cite{zingaleetal2009}. Other applications include  type I X-ray bursts
\cite{malone_multidimensional_2011, malone_multidimensional_2014} and
convection in helium shells of sub-Chandrasekhar mass WDs in the
double-detonation progenitor model for SNIa \cite{zingale_low_2013,
  jacobs_low_2016}. In the simulations of the convective Urca process we present here, we use a velocity ``sponge'' to damp the velocities in the low-density gas above the star \cite{zingaleetal2009} so such regions do not restrict the time step. Also, in these simulations the electron chemical potential contribution to energy generation is included as a heating source after each Strang-split reaction integration but not as a heating source for temperature evolution during the reaction integration. Because MAESTRO calculates local temperature from density and either pressure or enthalpy, we expect this arrangement to have only a slight effect compared to the case of including the contribution to the temperature evolution. We will address this discrepancy with additional simulations.

\section{Results}

The simulations of the convective Urca process were performed in progenitor WDs near the Chandrasekhar mass. As the progenitor star simmers, the temperature in the core increases, which changes the extent of the convective zone. These preliminary simulations were meant to characterize the extent of the convective zone and allow us to determine the central temperatures and densities of the progenitor models that allow the convective zone to reach the \mrm{A=23} Urca shell. We show results from a simulation with central temperature \mrm{5.5\times 10^8}~K, central density \mrm{4.5\times 10^9}~\grampercc, and total mass of 1.395~\Msun.

We constructed hydrostatic models by choosing the central density, temperature, and species mass fractions and integrating the equation of hydrostatic equilibrium radially outward to obtain the initial spherical base state for \maestro. In each zone of the initial model, we solve for the mass fractions of \iso{Na}{23} and \iso{Ne}{23} such that the electron capture and beta decay contributions to \mrm{\dxiso{Na}{23}} and \mrm{\dxiso{Ne}{23}} summed to zero while maintaining a constant value of \mrm{\xiso{Na}{23}+\xiso{Ne}{23}}. This approach to setting the initial mass fractions ensured that the \mrm{A=23} Urca species were initially in equilibrium with each other in the hydrostatic background state. We also supplement the equation of hydrostatic equilibrium by requiring the initial model have an isentropic region interior to mass coordinate 0.5~\msun, with the outer region isothermal. This condition allows us to capture the effects of a core convection zone in the initial model thermodynamics. For the central conditions described above, we used a composition consisting of \mrm{\xiso{C}{12}=0.39975}, \mrm{\xiso{O}{16}=0.59975}, and \mrm{\xiso{Na}{23}+\xiso{Ne}{23}=5\times 10^{-4}}.

\begin{figure*}[!ht]
  \begin{minipage}{0.48\textwidth}
    \includegraphics[width=\linewidth]{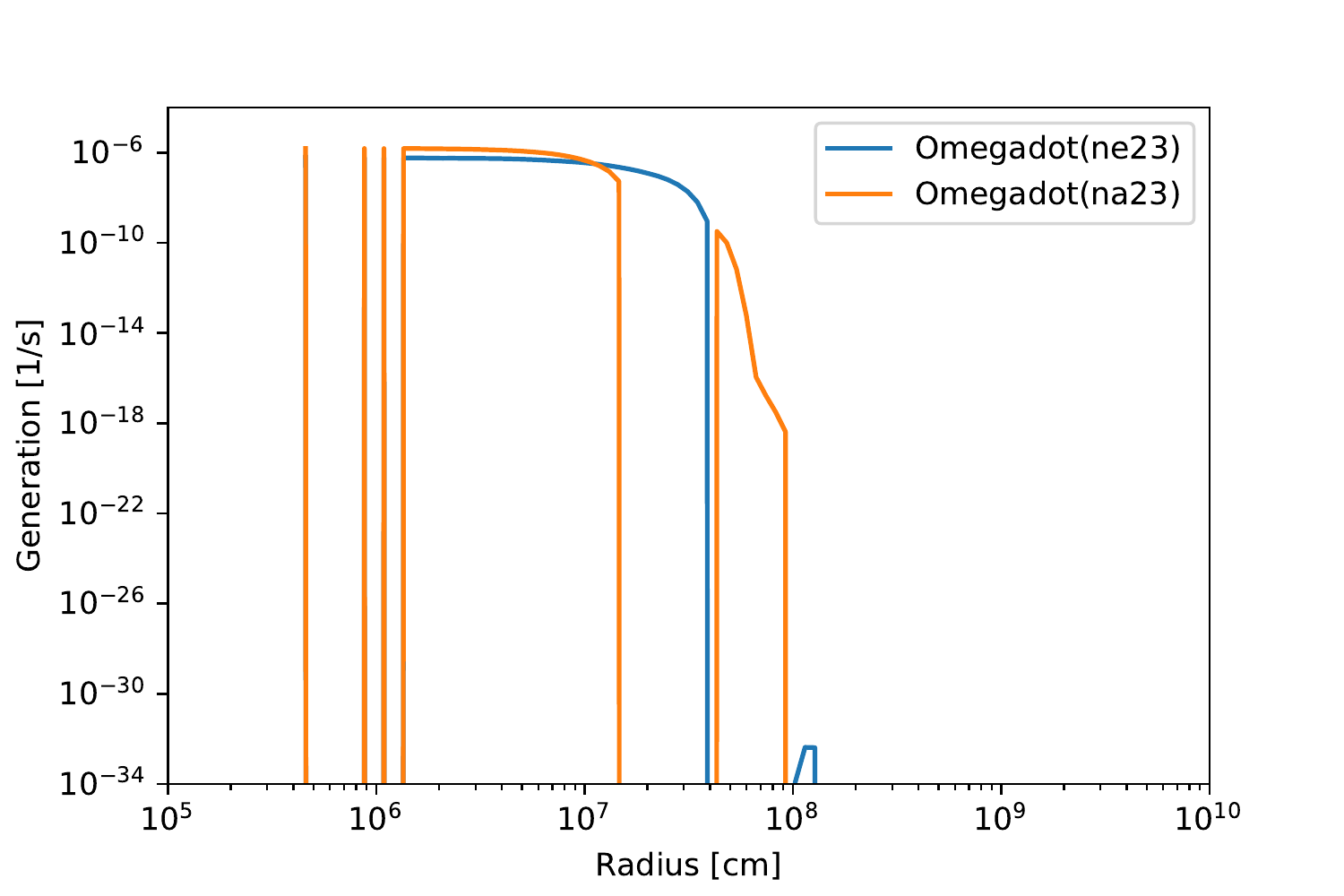}
  \end{minipage} \hfill
  \begin{minipage}{0.48\textwidth}
    \includegraphics[width=\linewidth]{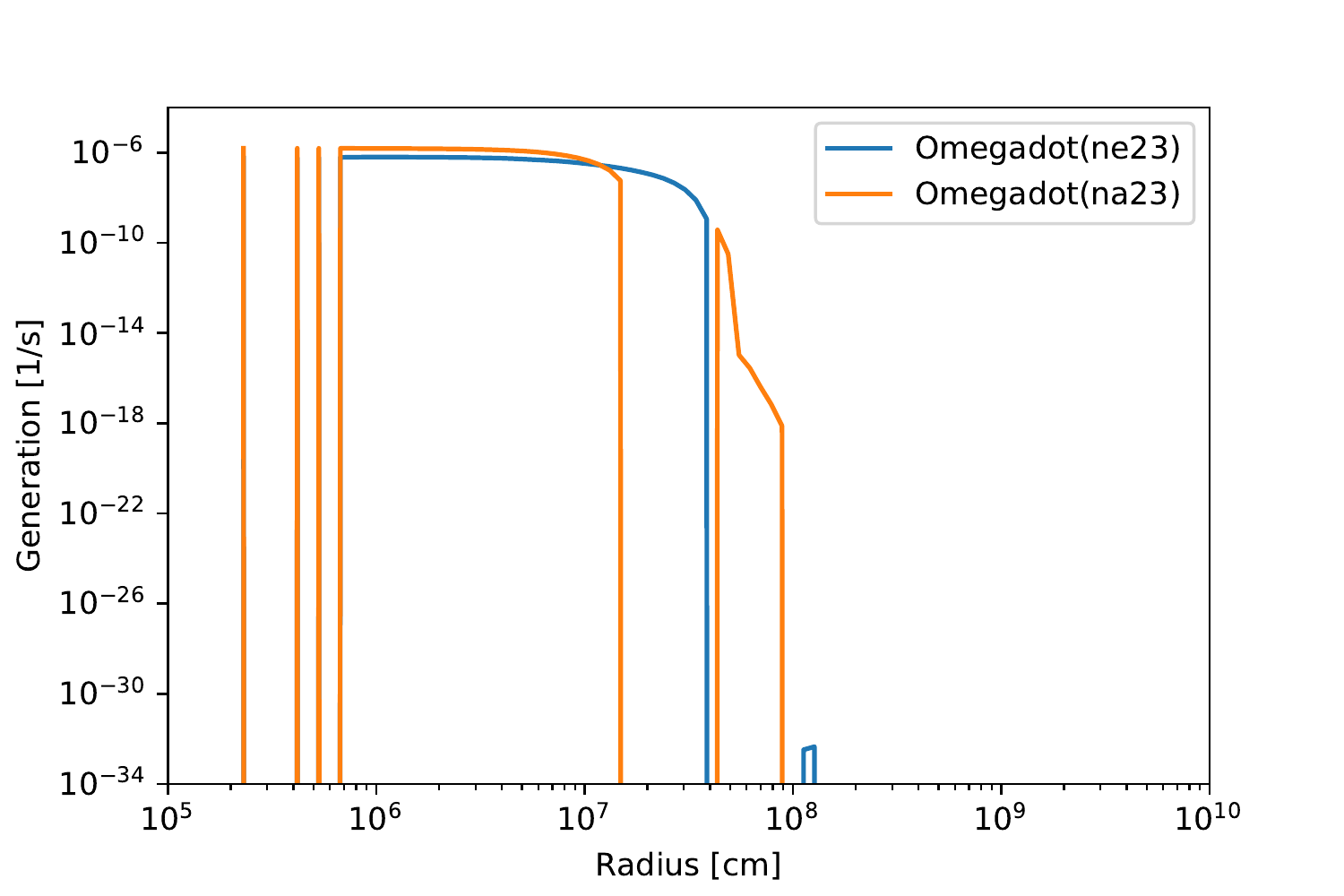}
  \end{minipage} \caption{\label{fig:omegadot_comparison} Plots of the production of $^{23}$Ne and $^{23}$Na within a white dwarf star as a function of radius. Left panel: Results at 3 levels of refinement, corresponding to an effective resolution of 5~\km~at a simulation time of 443.9 seconds. Right panel: Results at 4 levels of refinement, corresponding to an effective resolution of 2.5~\km~at a simulation time of 445.0 seconds.}
\end{figure*}

We initialized this model in \maestro\ using the multipole initialization method of \cite{zingaleetal2009} for the core velocity field and ran the two simulations at 5~\km\ and 2.5~\km\ effective resolution (3 and 4 spatial levels) to obtain the results shown here. We evolved these simulations for 842 and 627 seconds, respectively, corresponding to 5--10 convective eddy turnover times. In Figure \ref{fig:omegadot_comparison}, we show representative radially binned isotope production rates for \iso{Na}{23} and \iso{Ne}{23} in our two models, indicating radial zones for which \mrm{\dot{X}>0}. In these plots, we attribute the spikes at small radii to binning artifacts and note that at both resolutions the production rates for both \iso{Na}{23} and \iso{Ne}{23} are large within the core of the WD out to around 100~\km\ in radius, with \mrm{\dxiso{Na}{23}>\dxiso{Ne}{23}}. In this region within 100~\km, \iso{C}{12}-burning produces \iso{Na}{23}, which subsequently captures an electron to produce \iso{Ne}{23}. As the \iso{C}{12} burning rate drops between 100~\km\ and 200~\km\ in radius, we see the corresponding sharp decline in \iso{Na}{23} production. The region between 160~\km\ and 1000~\km\ is divided into two shells at 400~\km\ by the \mrm{A=23} Urca shell, where for \mrm{r < 400~\km}, electron captures onto \iso{Na}{23} produce \iso{Ne}{23}. The \iso{Na}{23} participating in this reaction is produced in the core from \iso{C}{12} fusion and transported into the core from above the Urca shell via convection. For the outer shell where \mrm{r > 400~\km}, \iso{Ne}{23} brought upwards by convection undergoes beta decay to \iso{Na}{23}. The bulk features presented in this radial binning appear qualitatively identical at the two resolutions, suggesting that the interaction between convection and reactions across the Urca shell is likely a physical effect and not an artifact of resolution in these simulations.

\begin{figure*}
\centering
  \begin{minipage}{\textwidth}
    \includegraphics[width=\linewidth]{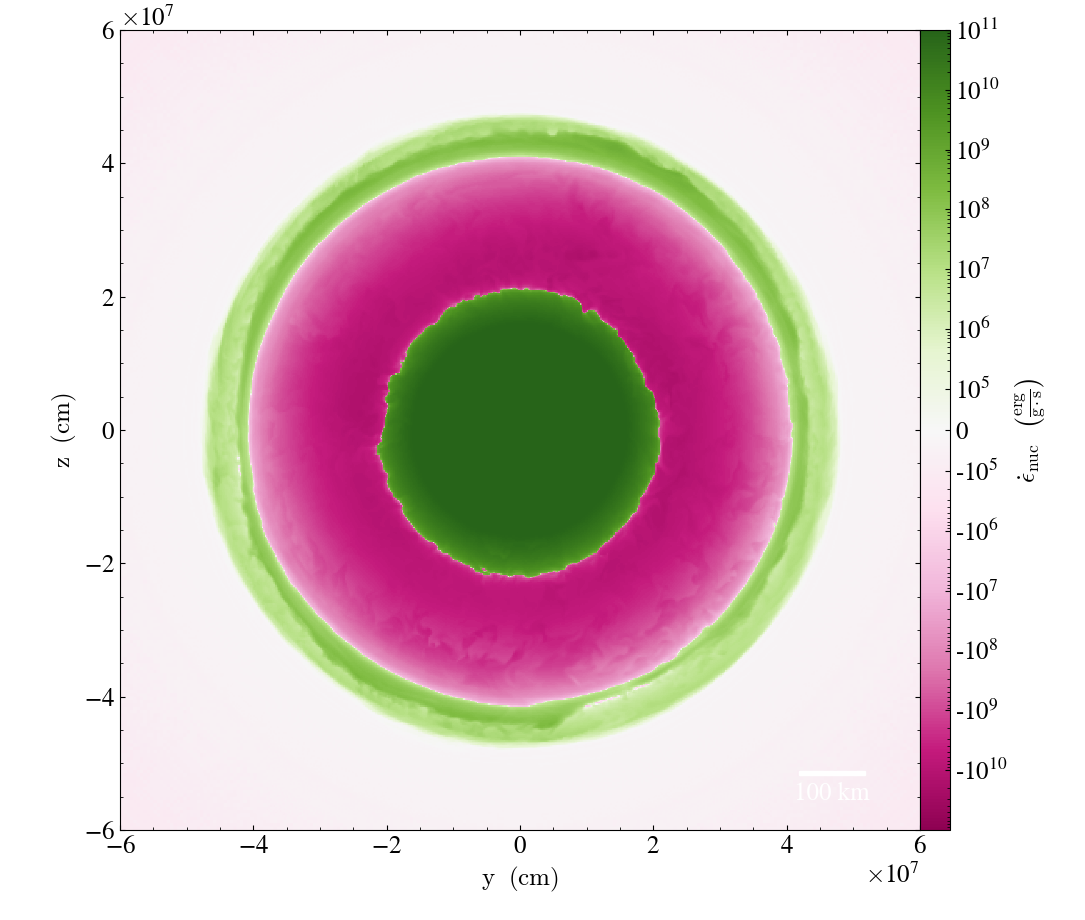}
  \end{minipage} \caption{\label{fig:enucdot_slice} Slice through the core of a WD at $x=0$ showing the specific energy generation rate as a result of \iso{C}{12} burning and the \mrm{A=23} Urca reactions. The WD shown has a central density of \mrm{4.5\times 10^9}~\grampercc, central temperature of \mrm{5.5\times 10^8}~K, and is spatially resolved to 2.5~\km\ in the energy generating regions shown. Positive specific energy generation is shown in green, and negative specific energy generation is shown in magenta.}
\end{figure*}

The energy generation structure arising from these reactions is shown in Figure \ref{fig:enucdot_slice} for the 2.5~\km-resolution model. The region of electron capture onto \iso{Na}{23} within approximately 400~\km\ in radius is evident and includes the central \iso{C}{12} burning region where the sum of reaction energy is exothermic. Between the exothermic burning core and the Urca shell lies an endothermic shell region where in the absence of burning electron captures occur either onto \iso{Na}{23} produced in the core and transported upwards or on \iso{Na}{23} transported downwards via convection from above the Urca shell. Outside the Urca shell the beta decay of \iso{Ne}{23} creates a thin exothermic shell structure at the upper edge of the convection zone. We note that while the bulk shell structure is spherical, the local energy generation rate can vary significantly depending on the local convective flow. This is to be expected from the convective Urca process, in which species transport across the Urca shell and the ensuing reactions are tightly coupled to convection.

\begin{figure*}[!ht]
  \begin{minipage}{0.48\textwidth}
    \includegraphics[width=\linewidth]{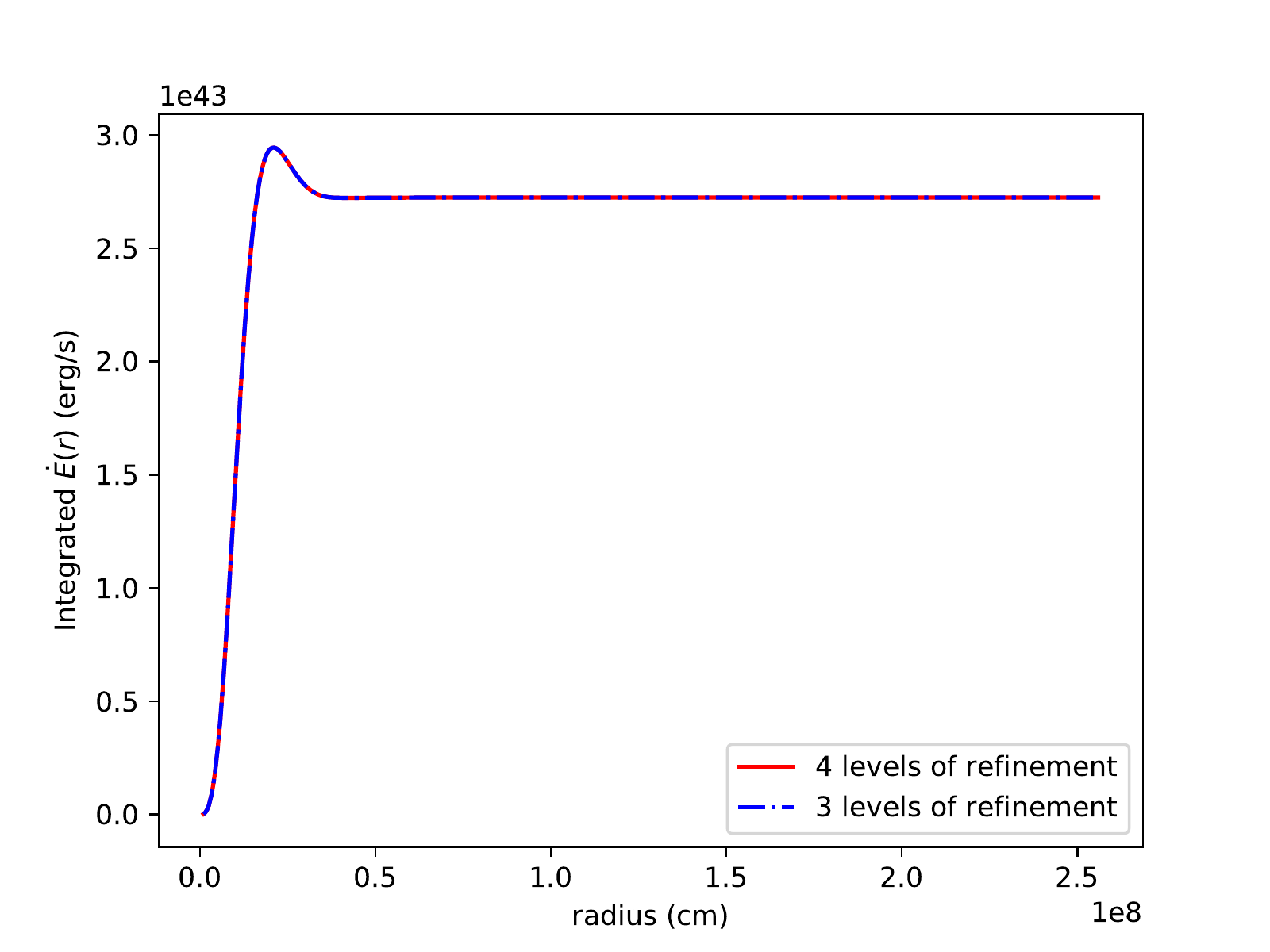}
  \end{minipage} \hfill
  \begin{minipage}{0.48\textwidth}
    \includegraphics[width=\linewidth]{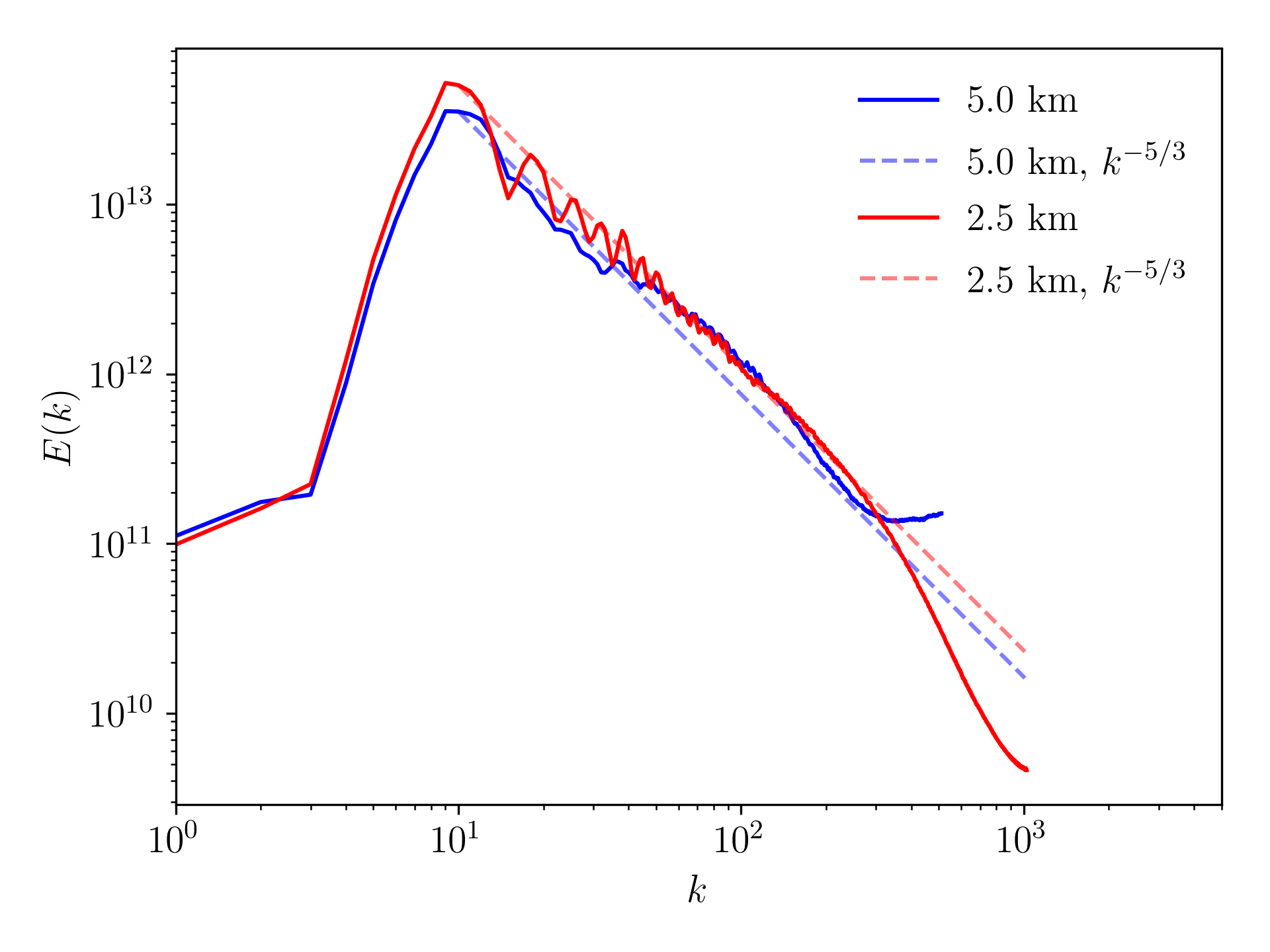}
  \end{minipage} \caption{\label{fig:edot_and_spectra} Left panel: Plot of integrated energy generation as a function of radius. Right panel: Plot of the turbulent kinetic energy power spectrum of the white dwarf. Red represents a simulation that was run with 4 levels of refinement corresponding to an effective resolution of 2.5~\km. Blue represents a simulation run with 3 levels of refinement corresponding to an effective resolution of 5~\km. The Kolmogorov scalings are plotted for these resolutions as dashed lines only to guide the eye and were matched to the simulation results at the tenth spectral point.}
\end{figure*}

%
%
%
The global energy generation arising from this shell structure is shown in Figure \ref{fig:edot_and_spectra} (left), where we plot the energy generation rate (in \ergspersecond) integrated from $r=0$ to $r=R$ as a function of radius $R$. The resolutions we explored yield consistent energy generation rates and we see that the peak energy generation integral occurs just outside the central \iso{C}{12} burning region and decreases due to the endothermic electron captures between the core and the Urca shell. Exothermic beta decays outside the Urca shell do not appear significant to the overall energy generation rate of the WD. We attribute this result to both the limited extent of the convection zone above the Urca shell and the energy loss to neutrinos during both the electron capture and beta decay reactions. We also note that the net effect of the Urca process under the conditions shown is to reduce the total energy generation slightly without having a net cooling effect on the WD, consistent with the results of \cite{Stein2006}.

We show in Figure \ref{fig:edot_and_spectra} (right) the turbulent kinetic energy power spectra computed for the models at 5~\km\ and 2.5~\km\ resolution at simulation times of 445.0 seconds and 443.9 seconds in blue and red, respectively. Over a large range of wavenumbers the 2.5~\km\ resolution simulation follows Kolmogorov scaling better than the coarser 5~\km\ simulation, although both simulations depart from Kolmogorov scaling at the largest wavenumbers. These kinetic energy power spectra were computed using a density weighting of $\rho^{1/3}$ for the spatial velocities and a cutoff density of $\mathrm{10^9~\grampercc}$ below which the velocity field is set to zero. The motivation for this is to compute the power spectra only over the portion of the domain corresponding to the convective WD core since the low density, non-convective regions of the WD exhibit spatial noise in the velocity field that may arise from unresolved gravity waves. For the 5~\km\ simulation, there is a small region between the convective core and the cutoff density containing such unphysical velocity noise that contributes to the excess energy at high wavenumber. In the 2.5~\km\ simulation, the corresponding region contains mostly gravity waves, which likely account for the energy deficit at high wavenumber. We are currently exploring this issue to understand better the resolution requirements of gravity waves in this model and the effect of unresolved gravity waves on the velocity field. We note that energy generation appears consistent between these two resolutions although convection is coupled to energy generation via the Urca process, suggesting that gravity waves outside the convective zone are not critical to the Urca process in this model. That will not in general be true for different central conditions and convection zone sizes, where it may be necessary for models with a smaller convection zone to resolve gravity waves above the Urca shell.

\section{Conclusions}

With the results presented here, we have working models for further investigation into
the convective Urca process. These models produce energy and react as we expect from previous studies and, importantly, we were able
to produce progenitors in which 
the convective region reaches the Urca shell, thus ensuring an
active convective Urca process. The models passed basic tests such as showing consistency between the
two resolutions even though the power spectra show the simulations may not fully capture the effects of 
turbulence at high wavenumbers, and the energy generation and cooling are consistent with earlier work. We conclude, therefore, that
these models are reasonable for further study of the convective Urca process.
Future work will include other pairs, e.g.\ \mrm{A=25}.

\ack

\maestro\ is freely available on GitHub
(\url{https://github.com/AMReX-Astro/}), and all problem setup files
for the calculations presented here are in the code repository.  This
work was supported in part by the US Department of Energy under grant
DE-FG02-87ER40317.  The reaction networks were generated using the \pynucastro\ library~\cite{pynucastro}.
Results in this paper were obtained using the
high-performance computing system at the Institute for Advanced
Computational Science at Stony Brook University. Support also came from the Data + Computing = Discovery summer program at
the Institute for Advanced Computational Science at Stony Brook University, supported in part by the NSF under a supplement to award AST-1211563. An award of computer
time was provided by the Innovative and Novel Computational Impact on
Theory and Experiment (INCITE) program.  This research used resources
of the Oak Ridge Leadership Computing Facility at the Oak Ridge
National Laboratory, which is supported by the Office of Science of
the U.S. Department of Energy under Contract No.\ DE-AC05-00OR22725.
This research used resources of the National Energy Research
Scientific Computing Center, which is supported by the Office of
Science of the U.S. Department of Energy under Contract
No.\ DE-AC02-05CH11231. Visualizations and part of this analysis made
use of \yt\ \cite{yt}. The authors thank Ilana Bromberg and Brianna Isola for previewing this manuscript.


\providecommand{\newblock}{}

\end{document}